\documentclass[showpacs,preprintnumbers]{revtex4}
\usepackage{amsfonts}
\usepackage{amssymb}
\usepackage{amsmath}
\usepackage{epsfig}
\usepackage{array}
\allowdisplaybreaks
\begin{document}
\title{Spatially Hyperbolic Gravitating Sources in $\Lambda$-Dominated Era}
\author{Z. Yousaf}
\email{zeeshan.math@pu.edu.pk}
\affiliation{Department of Mathematics, University of the Punjab, Quaid-i-Azam Campus, Lahore-54590, Pakistan.}


\begin{abstract}
This study focuses on the impact of cosmological constant on hyperbolically symmetric matter configurations in static background. We extend the work of Herrera \cite{herrera2021hyperbolically} and describe the influences of such repulsive character on a few realistic features of hyperbolical anisotropic fluids. After describing Einstein-$\Lambda$ equations of motion, we elaborated the corresponding mass function along with its conservation laws. In our study, besides observing negative energy density, we notice the formation of Minkowskian core as matter content is compelled itself not to follow inward motion near the axis of symmetry. Three families of solution are found in $\Lambda$-dominated epoch. First is calculated by keeping zero value to Weyl scalar, while the second solution maintains zero complexity in the subsequent changes of the hyperbolical compact object. However, the last model encompasses stiff fluid within  the self-gravitating system. Such type of theoretical setup suggests its direct link to study a few particular quantum scenarios where negative behavior of energy density is noticed at $\Lambda$ dominated regime.
\end{abstract}
\maketitle

\section{Introduction}

General relativity (GR) is a very remarkable and astonishing theory that has resolved many issues since its birth. It is considered as a pillar of modern physics and has revolutionized the understanding of the universe. The gravitational red shift, precession of perihelion of Mercury, and light bending by the Sun are predicted by GR. Although there is no exaggeration to say that GR is the most successful theory, yet many issues are still open. Some of the important unresolved issues are the localization of energy and singularity problem, etc. In addition, the most fundamental and fascinating theoretical problem of 21st century is considered to be current accelerating expansion of the universe or dark energy problem. Initially, Einstein believed that the universe is neither expanding nor contracting but static. In order to show this behavior of the universe, Einstein modified the field equations by adding a cosmological constant $``\Lambda"$. Later on (1929), Hubble performed an experiment on more than twenty galaxies and observed that light coming from these galaxies is red shifted which shows that the universe is expanding. After this, Einstein admitted that his idea of static universe was the biggest blunder of his life. So he removed $\Lambda$ from the field equations.

Recent observations of astrophysics and modern cosmology like cosmic microwave background radiations, Supernovae surveys, and large scale structures of our universe \cite{pietrobon2006integrated,riess2007new,tegmark2004cosmological,giannantonio2006high} indicate that at present times, the universe is expanding with acceleration. The energy composition of the universe turns out to be 4.9\% of ordinary baryonic matter, 26.8\% of dark matter (DM) and 68.3\% of dark energy (DE). The description provided by GR may need to modify to understand this observational data. There are several approaches to describe the cosmic acceleration out of which the inclusion of cosmological constant in Einstein field equations is one of the the primary models used for this purpose. The term DM means a form of matter which is unknown and cannot be deducted by its radiation but the gravitational force is exerted by it. Dark energy refers as unknown energy and an exotic substance having large negative pressure. The cosmological constant is the primary and most relevant candidate of DE which provides the most reasonable explanation of the universe. The notion of $\Lambda$ could be used to explain inflationary era in the early universe.

Bowers and Liang \cite{bowers1974anisotropic} established a framework to study effect of anisotropy on self-gravitating stars by the observer. Herrera and Santos \cite{herrera1997local} considered the radial perturbation approach to analyse a few properties of anisotropic self-gravitating spheres. Di Prisco \emph{et al.} \cite{di1997cracking} found little but continuous variation of pressure anisotropy during the phenomenon of cracking of the relativistic celestial spherical bodies. Abreu \emph{et al.} \cite{abreu2007sound} checked the impact of local anisotropy on the existence of matter distributions in GR after being its departure from the state of equilibrium. They also
analyzed the stability of star with the help of tangential and radial sound speeds. Bhar \emph{et al.} \cite{bhar2016modelling} performed theoretical analysis on static diagonally symmetric relativistic spheres in GR and found few stable stellar models due to anisotropic pressure. Maurya and Govender \cite{maurya2017family} analyzed physical analysis of their calculated exact solutions in the formation of strange anisotropic stars. They found anisotropic pressure as an additional force in  sustaining the star stability against cracking.

Morales and Tello-Ortiz \cite{morales2018charged} calculated few anisotropic solutions for the relativistic spherical structures. After performing stability tests, they inferred that their anisotropic solutions are well behaved and well-posed. Chanda \emph{et al.} \cite{chanda2019anisotropic} found that static irrotational stellar models can be described well if such systems have a high level of anisotropic stresses around the central region. Yousaf \emph{et al.} \cite{yousaf2019non,sharif2015instability,10.1093/mnras/staa1470,yousaf2021quasi} performed mathematical modeling of radiating and non-radiating matter configurations in an environment of unequal principal stresses and found stable fluid configurations under specific parametric choices. Raposo \emph{et al.} \cite{raposo2019anisotropic} described the modeling of anisotropic self-gravitating objects and found relatively more massive and compact objects due to the presence of anisotropicity in pressure. Godani and Samanta \cite{samanta2019wormhole,godani2020traversable} found stable anisotropic wormhole solutions in modified gravity through graphical representations. G\'{o}mez-Leyton \emph{et al.} \cite{gomez2020charged} explored stable exact model of locally anisotropic matter configurations. After computing a relation among anisotropic pressure and matter variables, they found stable regimes for spherical stars.

Hyperbolically symmetrical spacetimes (HSS) are thought to be capable of explaining many unresolved cosmological puzzles. Harrison \cite{harrison1959exact} considered HSS and obtained the corresponding analytical models through separation of variable method. Malik \emph{et al.} \cite{malik2015mhd} described the distribution of hyperbolically symmetric matter over the cylindrical geometry and presented a few numerical solutions after solving the corresponding differential equations. Gaudin \emph{et al.} \cite{gaudin2006gravity} calculated exact analytical models of GR equation of motion with massless scalar space and described a few characteristics of hyperbolic metric in vacuum. Maciel \emph{et al.} \cite{maciel2020new} explored locally isotropic solutions of interior spacetime and claimed that such incompressible spherical solution could be treated as a unique with HSS background. Herrera \emph{et al.} \cite{herrera2021hyperbolically} presented a HSS version of LTB geometry and explored its less complex exact solutions. Miguel \cite{sanchez2021some} studied few characteristics of HSS and found the non-convex Cauchy temporal function for the conformal families of solutions. Cao and Wu \cite{cao2021} considered the metric version of $f(R)$ gravity and discussed the occurrence of strong hyperbolicity under some specific gauge constraints. Recently, Herrera \emph{et al.} \cite{herrera2021hyperbolically} carried out a brief study on HSS after calculating conformally flat and zero complexity solutions. They also described the physical applications by adopting a general approach in solving analytical solutions. Bhatti \emph{et el.} \cite{bhatti2021influence} described conformally flat solutions of HSS in static charged medium. Yousaf \emph{et al.} \cite{yousaf2022hyper} calculated the Tolman mass of HSS and then explored the occurrence of core formation by taking constant energy density of the matter distribution.

Structure scalars are the scalars which are obtained from the orthogonal splitting of the Riemann tensor introduced by Bel \cite{bel1961inductions}. They are related to the fundamental properties of the fluid directly. Timelike vectors are useful in order to perform orthogonal splitting of a tensor, which gives rise to three tensors, namely $X_{\mu\nu},~Y_{\mu\nu}$ and $Z_{\mu\nu}$. The scalars corresponding to these, i.e., $Y_T,~X_T,~Z,~Y_{TF}$ and $X_{TF}$ are called structure scalars \cite{herrera2009structure}. Here $Y_T$ appears to be proportional to the Tolman mass density for systems in equilibrium \cite{herrera2012cylindrically,yousaf2016electromagnetic}. It also controls the evolution of expansion scalar, $X_T$ gives energy density of the fluid and $Z$ deals with dissipation fluxes for non-spherical systems. Because of these scalars, it is easy to deal with a complex systems (as compared to vectors and tensors). This single tool deals with different aspects of the system providing a lot of information about evolution of the system, e.g., shear and expansion evolution, inhomogeneity, complexity, etc can be dealt with these structure scalars \cite{herrera2011role,herrera2010lemaitre,herrera2014dissipative}. Yousaf and his collaborators \cite{yousaf2016causes,yousaf2016influence,doi:10.1142/S0217732319503334,yousaf2020definition,bhatti2021electromagnetic,yousaf2021quasi} computed the modified version of these scalars by invoking extra curvature terms and described their influences in understanding evolutionary mechanism of adiabatic and non-adiabatic relativistic structures. Herrera \cite{herrera2018new} introduced a new concept of complexity factor in terms of one of these scalars for the static anisotropic spherical objects. He concluded that $Y_{TF}$ is playing the role in the fixing of complexity of relativistic system. The applications of these results are then provided for various cosmic and stellar backgrounds by first Herrera \emph{et al. } \cite{herrera2018definition,herrera2019complexitya,herrera2019complexity} and then by their followers \cite{yousaf2020complexity,bhatti2021structure,yousaf2021measure}.

This paper is aimed to study the effects of cosmological constant on HSS analysis by following the procedure initially presented by Herrera \emph{et al.} \cite{herrera2021hyperbolically}.
After describing the metric as well as its gravitating source, we shall describe total matter quantity of the HSS through two well-known formalisms at $\Lambda$-dominated epoch in section \textbf{2}. Few important relations among these with the matter variables will also be specified in the same section. Section \textbf{3} is devoted to the calculation of structure scalars from the decomposition of the curvature tensor. Three different families of exact analytical models will be established for Einstein-$\Lambda$ gravity in section \textbf{4}. The last section will summarize our results and discussions.

\section{Spatially Hyperbolic Geometry and Matter Content}

We assume two different types of boundary surfaces, i.e., external and internal ones. We shall represent these with $\Omega^e$ and $\Omega^i$, respectively. The hypersurface $\Omega^e$
demarcates the static anisotropic fluid with the external de-Sitter vacuum spacetime, while $\Omega^i$ keeps the matter configuration from the central vacuole. This type of mathematical model gives rise to the formation of central Minkowskian cavity. Such types of cosmic configurations could be effective to discuss cosmological voids and other aspects of mathematical cosmology.

We take the following form of line element as
\begin{equation}\label{h5}
ds^2=e^{\lambda(r)}dt^2-e^{\nu(r)}dr^2-r^2d\omega^2,
\end{equation}
where $d\omega^2=d\theta^2-r^2 \sinh^2 \theta d\phi^2$, and metric variables are dependent on $r$ only. We assume the Eckart frame, this would keep the system's fluid to be in a rest state. The above system is assumed to be coupled with the locally anisotropic fluid which can be written mathematically as
\begin{equation}\label{h7}
T_{\nu\mu}=(\mu+P)V_\nu V_\mu-Pg_{\nu\mu}+\Pi_{\nu\mu},
\end{equation}
where $\mu$ is the eigen value of energy momentum tensor with respect to the eigen vector $V_\nu$, $P$ is the pressure and $\Pi_{\nu\mu}$ is the anisotropic tensor. These can be defined through projection tensor $h_{\nu\mu}$ as
$$\Pi_{\nu\mu}=h^\beta_{~\nu}h^{\alpha}_{~\mu}(T_{\beta\alpha}+Ph_{\beta\alpha}), \quad P=-1/3\times T_{\nu\mu}h^{\nu\mu}.$$
We take the following vacuum solution of the field equation outside $\Omega^e$ with de-Sitter Schwarzschild hyperbolic spacetime as
\begin{equation}\label{h3}
ds^2=\left(\frac{\Lambda C^2}{3}+\frac{2M}{C}-1\right)dt^2-\left(\frac{\Lambda C^2}{3}+\frac{2M}{C}-1\right)^{-1}dC^2-C^2 d\theta^2-C^2 \sinh^2\theta d\phi^2,
\end{equation}
where $\Lambda,~C$ and $M$ are the cosmological constant, radial distance and constant mass, respectively. We are interested to perform our analysis in the region
$6M>3C-\Lambda C^3$. It is worthy to note that the space inner to the black hole horizon can be well discussed through the above metric. However, the outer manifold to black hole
horizon can be described with the usual de-Sitter Schwarzschild metric. The matching of the above metric with the interior fluid requires the fulfilment of the Darmois junction conditions \cite{darmois1927memorial}. These conditions facilitate the smooth joining of the both exterior and interior manifolds at $\Omega^e$. These constraints after calculations are found as under
\begin{align}\label{h25}
e^{\lambda}\overset{\Omega^{(e)}}=\frac{\Lambda C^2}{3}+\frac{2M}{C}-1,\quad e^{\nu}\overset{\Omega^{(e)}}=\frac{1}{\frac{\Lambda C^2}{3}+\frac{2M}{C}-1},\quad
P_{xx}\overset{\Omega^{(e)}}=\Lambda.
\end{align}
Next, we adopt Bondi approach \cite{bondi1964proceedings} to get HSS from the generic formulation of the axial static symmetric manifold. We define the locally Minkowskian frame (LMF)
associated with tetrad field as
\begin{equation}\nonumber
d\breve{\tau}=e^{\nu/2}dt; \quad d\breve{x}=e^{\lambda/2}dr; \quad d\breve{y}=rd\theta; \quad d\breve{z}=r \sinh\theta d\phi.
\end{equation}
Here, we use the notation breve indicates that the terms are estimated by an observer residing in LMF. Then the associated stress-energy tensor is described through principal stresses $P_{yy},~P_{xx},~P_{zz},~P_{yx}$ and $\mu$ as follows
$$ \breve{T}_{\nu\mu}=
\begin{pmatrix}
\mu & 0 & 0 & 0\\
0 & P_{xx} & P_{xy} & 0\\
0 & P_{yx} & P_{yy} & 0\\
0 & 0 & 0 & P_{zz} \\
\end{pmatrix}.
$$
Such kind of configuration is found to produce axially symmetric gravitational geometry. It is noticed that the HSS experienced  $P_{xx}\neq P_{yy}=P_{zz}$ along with the null contribution of $P_{xy}$. In an environment of Minkowskian frame, the fours vector are defined as under
\begin{align}\nonumber
\breve{V}_\eta=(1,0,0,0), \quad \breve{K}_\eta=(0,-1,0,0), \quad \breve{L}_\eta=(0,0,-1,0), \quad \breve{S}_\eta=(0,0,0,-1).
\end{align}
With the help of these vectors, one can elaborate fluid distribution mathematically as
\begin{equation}\nonumber
\breve{T}_{\nu\mu}=(\mu+P_{zz})\breve{V}_\nu \breve{V}_\mu-P_{zz}\eta_{\nu\mu}+(P_{xx}-P_{zz})\breve{K}_\nu\breve{K}_\mu.
\end{equation}
Thus the back transformation to $(t,r,\theta,\phi)$ in LMF provides the following form of matter content for HSS as
\begin{equation}\label{h18}
T_{\alpha\xi}=(\mu+P_{zz})V_\alpha V_\xi-P_{zz}g_{\alpha\xi}+(P_{xx}-P_{zz})K_\nu{K}_\mu,
\end{equation}
where $V_\xi=(e^{\nu/2},0,0,0),~K_\eta=(0,-e^{\lambda/2},0,0)$ in the comoving reference frame and $\Pi=P_{xx}-P_{zz}$.
The smooth joining of locally anisotropic HSS matter content with the Minkowski metric at the internal boundary can be
dealt with Darmois conditions \cite{darmois1927memorial}. These are found after few calculation as under
\begin{equation}\label{h26}
e^{\nu}\overset{\Omega^{(i)}}{=}1,\quad e^{\lambda}\overset{\Omega^{(i)}}{=}1,\quad m(t,r)\overset{\Omega^{(i)}}{=}\frac{\Lambda C^3}{6},\quad
P_{r}\overset{\Omega^{(i)}}{=}\Lambda.
\end{equation}
Here, the matter content is compelled itself not to follow inward motion near the axis of symmetry. This gives rise to develop Minkowskian core.

Now, the equations of motion for Einstein-$\Lambda$ gravity are described as
\begin{equation}\label{h6}
G_{\nu\mu}\equiv
R_{\nu\mu}-\frac{1}{2}Rg_{\nu\mu}=\kappa
T_{\nu\mu}-\Lambda g_{\nu\mu},
\end{equation}
where $R,~g_{\gamma\delta},~R_{\gamma\delta}$ are the Ricci scalar, gravitational potential, and the Ricci
tensors, respectively. These equations for the set of system \eqref{h5} and \eqref{h18} turn out to be
\begin{align}\label{h27}
8\pi\left({\rho}+\frac{\Lambda}{8\pi}\right)&=-\frac{1+e^{-\nu}}{r^2}+\frac{\nu'e^{-\nu}}{r},\\\label{h28}
8\pi\left(P_{xx}-\frac{\Lambda}{8\pi}\right)&=\frac{1+e^{-\nu}}{r^2}+\frac{\lambda'e^{-\nu}}{r},\\\label{h29}
8\pi\left(P_{yy}-\frac{\Lambda}{8\pi}\right)&=\frac{e^{-\nu}}{2}\left(\lambda''-\frac{\lambda'\nu'}{2}+\frac{\lambda'^2}{2}
-\frac{\nu'}{r}+\frac{\lambda'}{r}\right).
\end{align}
We shall use $P_{xx}=P_r$ and $P_{yy}=P_{zz}=P_\bot$ in the coming equations. The law of conservation at $\Lambda$-dominated regime for the static locally anisotropic
relativistic spherical interiors are found as under
\begin{equation}\label{h30}
P_r'+\frac{\lambda'}{2}(\mu+P_r)+\frac{2\Pi}{r}=0.
\end{equation}

The Misner-Sharp formalism \cite{PhysRev.136.B571} provides the following configurations of the mass function
\begin{equation}\label{h32}
m(r)=\frac{r}{2}\left(1+e^{-\nu}\right),
\end{equation}
which can be rendered after using field equation as
\begin{equation}\label{h33}
m(r)=-4\pi \int_0^r\left(\mu+\frac{\Lambda}{4\pi}\right)r^2dr.
\end{equation}
It would be interesting to notice from Eq.\eqref{h32} that mass of the spatially hyperbolic
object is positive. In order to maintain this logic, one can analyze from the above equation that
energy density should be negative. This leads to the breaching of weak energy conditions by the HSS.
To avoid negativity of energy density, we shall use $\mu$ instead of $-|\mu|$ in our calculations. It follows from Eq.\eqref{h33} that
\begin{equation}\nonumber
m(r)=4\pi\int_{r_{min}}^r\left|\mu+\frac{\Lambda}{4\pi} \right|r^2dr.
\end{equation}
Equations \eqref{h28} and \eqref{h32} give
\begin{equation}\label{h35}
\lambda'=2\left\{\frac{4\pi \left(P_r -\frac{\Lambda}{8\pi}\right)r^3-m}{r(2m-r)}\right\}.
\end{equation}
Upon making use of the above value, we get from Eq.\eqref{h30}
\begin{equation}\label{h36}
P_r'+\left(\frac{4\pi \left(P_r -\frac{\Lambda}{8\pi}\right)r^3-m}{r(2m-r)}\right)(P_r-|\mu|)+\frac{2\Pi}{r}=0.
\end{equation}
This equation describes the state of hydrostatic equilibrium of the HSS within Einstein-$\Lambda$ gravity.

\section{Active Gravitating Mass with Cosmological Constant}

The well-known Weyl tensor can be written through its electric part, fluid four velocity and Levi-Civita tensor ($\eta_{\pi\lambda\gamma\delta}$) as
\begin{equation}\nonumber
C_{\xi\nu\pi\lambda}=E^{\beta\delta}V^\rho V^\gamma
(g_{\xi\nu\rho\beta}
g_{\pi\lambda\gamma\delta}-\eta_{\xi\nu\rho\beta}
\eta_{\pi\lambda\gamma\delta}),
\end{equation}
where
$g_{\xi\nu\rho\beta}=g_{\xi\rho}g_{\nu\beta}-g_{\xi\beta}g_{\nu\rho}$. The scalar associated with the electric part of the above equation is calculated for the static spheres as
\begin{align}\label{h41}
\mathcal{E}&=-\frac{{\lambda}''e^{-\nu}}{4}-\frac{\lambda'^2
e^{-\nu}}{8}+\frac{\nu'\lambda'e^{-\nu}}{8}+\frac{\nu'e^{-\nu}}{4r}-
\frac{\nu'e^{-\nu}}{4r}-\frac{e^{-\nu}}{2r^2}-\frac{1}{2r^2}.
\end{align}
This after making use of equations of motion, Eqs.\eqref{h32} and \eqref{h41} give the following form of the Misner-Sharp mass function
\begin{equation}\label{h42}
\frac{3m}{r^3}=4\pi\left|\mu+\frac{\Lambda}{4\pi}\right|+4\pi\Pi-\mathcal{E}.
\end{equation}
Equation (\ref{h33}) after using above expression provides
\begin{equation}\label{h43}
\mathcal{E}=4\pi\Pi+\frac{4\pi}{r^3}\int_0^r \left|\mu+\frac{\Lambda}{4\pi}\right|'r^3 dr.
\end{equation}
Feeding back $\mathcal{E}$ from the above equation in Eq.\eqref{h42} gives
\begin{equation}\label{h44}
m=\left|\mu+\frac{\Lambda}{4\pi}\right|\times\frac{4\pi r^3}{3}-\frac{4\pi}{3}\int_0^r \left|\mu+\frac{\Lambda}{4\pi}\right|'r^3 dr.
\end{equation}
In this way, we have been able to relate the Misner-Sharp function with the system energy density and cosmological constant. One can realize the effects of dark energy effects through $\Lambda$ in the static fluids from the above expression.

Now, we adopt another approach to calculate the quantity of matter ingredients of the static relativistic compact bodies. Since, in HSS, the two types of boundary surfaces are likely to appear. Therefore, at the outer surface, we can provide the generic formula for finding the active gravitating matter quantity for HSS in an environment of cosmological constant as
\begin{equation}\nonumber
m_T=\int^{2\pi}_0 \int^\pi_0 \int^r_0 r^2 \sinh \theta e^\frac{\nu+\lambda}{2}(T^0_0-T^1_1-2T^2_2)d\tilde{r}d\theta d\phi,
\end{equation}
which after using equations of motion at $\Lambda$-dominated epoch give
\begin{equation}\label{h45}
m_T=2\pi(cosh\pi-1)\int^r_0 e^\frac{\nu+\lambda}{2}\tilde{r}^2 \left(-|\mu|+P_r-\frac{2\Lambda}{\kappa}+2P_\bot\right)d\tilde{r}.
\end{equation}
Its solution is found as under
\begin{equation}\label{h46}
m_T=\frac{\cosh\pi-1}{4}\lambda'r^2 e^\frac{\lambda-\nu}{2}.
\end{equation}
After substituting the value of the metric co-efficient from Eq.\eqref{h35}, it follows that
\begin{equation}\label{h47}
m_T=\frac{\cosh\pi-1}{2}\left[4\pi \left(P_r-\frac{\Lambda}{4\pi}\right)r^3-m\right]e^\frac{\nu+\lambda}{2},
\end{equation}
thereby providing the repulsive nature (if $4\pi P_r r^3-m<\Lambda r^3$) induced by the gravitational force within the locally anisotropic static metric. This also described the importance of
$\Lambda$ terms in the estimation of active gravitating static matter. Now, we wish to relate this mass function in terms of scalar corresponding to four acceleration ($a_\nu$). One can easily check $a_\nu=aK_\nu$ with $a=\frac{\lambda' e^\frac{-\nu}{2}}{2}$. This can be reexpressed as
\begin{equation}\nonumber
a=\frac{2 e^\frac{-\lambda}{2}m_T}{r^2 (\cosh\pi-1)}.
\end{equation}
The simultaneous use of Eqs.(\ref{h45}) and \eqref{h47} provide
\begin{equation}\nonumber
m'_T-\frac{3m_T}{r}= -\left({\frac{\cosh\pi-1}{2}}\right)r^2 e^\frac{\nu+\lambda}{2}\left(\mathcal{E}+4\pi\Pi\right).
\end{equation}
This is first order differential equation in $m_T$. Its solution is found as under
\begin{equation}\nonumber
m_T=\frac{r^3_{\Omega_e}(m_T)}{r^3_{\Omega^e}}+\left(\frac{\cosh\pi-1}{2}\right)r^3 \int^{r_{\Omega^e}}_r
\left(\mathcal{E}+4\pi\Pi\right)\frac{e^\frac{\nu+\lambda}{2}}{\tilde{r}}d\tilde{r}.
\end{equation}
This provides the expression of Tolman mass in terms of cosmological constant, tidal forces, and locally anisotropic pressure. The above equation after substituting
$\mathcal{E}$ yields
\begin{equation}\label{h54}
m_T=\frac{r^3_{\Omega_e}(m_T)}{r^3_{\Omega^e}}+\left(\frac{\cosh\pi-1}{2}\right)r^3 \int^{r_{\Omega^e}}_r \left[8\pi\Pi+\frac{4\pi}{\tilde{r}^3}
\int^r_0 \left|\mu+\frac{\Lambda}{4\pi}\right|' \tilde{r}^3 d\tilde{r}\right]\frac{e^\frac{\nu+\lambda}{2}}
{\tilde{r}}d\tilde{r}.
\end{equation}
This relation has related Tolman spherical mass with the inhomogeneous energy density with repulsive energy effects due to $\Lambda$ terms and local anisotropic pressure. This result reduces to GR upon substituting $\Lambda=0$.

Herrera \emph{et al.} \cite{herrera2009structure,Herrera2012} described the orthogonal decomposition of the curvature tensor into
two different tensorial objects. These can be found for our system in the background of cosmological constant as under
\begin{align}\label{h55}
&X_{\alpha\beta}=~^{*}R^{*}_{\alpha\gamma\beta\delta}V^{\gamma}V^{\delta}=
\frac{1}{2}\eta^{\varepsilon\rho}_{~~\alpha\gamma}R^{*}_{\epsilon
\rho\beta\delta}V^{\gamma}V^{\delta},\\\label{h57}
&Y_{\alpha\beta}=R_{\alpha\gamma\beta\delta}V^{\gamma}V^{\delta},
\end{align}
where $R^{*}_{\alpha\beta\gamma\delta},~^{*}R_{\alpha\beta\gamma\delta}$ stand for the right and left dual of curvature object. The trace (T) and trace-less (TF) values of the above equations after using field equations become
\begin{align}\label{h70}
X_T&=-8\pi\left|\mu+\frac{\Lambda}{4\pi}\right|,\quad
X_{TF}=4\pi\Pi-\mathcal{E},\\\label{h72}
Y_T&=4\pi(-\left|\mu+\frac{\Lambda}{4\pi}\right|+3P),\quad
Y_{TF}=4\pi\Pi+\mathcal{E}.
\end{align}
The second of Eqs.\eqref{h70} and \eqref{h72} after using $\mathcal{E}$ from Eq.\eqref{h43} turn out to be
\begin{align}\label{h71}
X_{TF}&=-\frac{4\pi}{r^3}\int^r_0 \tilde{r}^3\left|\mu+\frac{\Lambda}{4\pi}\right|'d\tilde{r},\\\label{h74}
Y_{TF}&=8\pi\Pi+\frac{4\pi}{r^3}\int^r_0 \tilde{r}^3\left|\mu+\frac{\Lambda}{4\pi}\right|'d\tilde{r}.
\end{align}
One can notice that $X_{TF}$ is trying to control the effects of energy density inhomogeneity of the self-gravitating system in Einstein-$\Lambda$ gravity,
while the $Y_{TF}$ not only takes the effects of energy density but also the influences of pressure anisotropicity on the subsequent changes within the static spherical bodies.
The effects of pressure can be dealt through trace less parts of Eqs.\eqref{h55} and \eqref{h57} as
\begin{equation}\label{h75}
X_{TF}+Y_{TF}=8\pi\Pi.
\end{equation}
Equations \eqref{h45} and \eqref{h54} turn the trace and trace-less components as follows
\begin{align}\label{h76}
m_T&=\frac{r^3_{\Omega_e}(m_T)}{r^3_{\Omega^e}}+\left(\frac{\cosh\pi-1}{2}\right)r^3\int^{r_{\Omega e}}_r
\frac{Y_{TF}}{s}e^\frac{\nu+\lambda}{2}ds,\\\label{h77}
m_T&=\frac{\cosh\pi-1}{2}\int_0^r s^2 Y_T e^{(\nu+\lambda)/2}ds.
\end{align}
The first of the above equation provides us a way to study the non-complex state of the system through Tolman mass. This equation has linked $Y_{TF}$ with $m_T$. Thus the combined
analysis of Eqs.\eqref{h74} and \eqref{h76} states that $Y_{TF}$ could be treated as a complexity factor. The second of above equation describes the direct connection of $Y_T$ with gravitating passive static fluid mass of the relativistic system even in the presence of $\Lambda$. This result is equivalent to one obtained by Herrera \emph{et al.} \cite{herrera2021hyperbolically} in the absence of $\Lambda$.

\section{Static HSS Fluids}

In this section, we present an analytical solution of Einstein-$\Lambda$ equations of motion for the anisotropic HSS by following a general method presented by Lake \cite{lake2004galactic} and Herrera \emph{et al.} \cite{herrera2008all}. The subtraction of Eq.\eqref{h29} from Eq.\eqref{h28} provides
\begin{equation}\label{h78}
8\pi(P_r-P_\bot)=\frac{1+e^{-\nu}}{r^2}-\frac{e^{-\nu}}{2}\left(\lambda''+\frac{\lambda'^2}{2}-\frac{\lambda'\nu'}{2}
-\frac{\lambda'}{r}-\frac{\nu'}{r}\right).
\end{equation}
To proceed forward for the sake of solutions, we define $\frac{\lambda'}{2}\equiv \mathfrak{z}-\frac{1}{r}$ and $\mathfrak{a}\equiv e^{-\nu}$. After this, above equations turns out to be
\begin{align}\nonumber
\mathfrak{a}'+\mathfrak{a}\left[\frac{4}{r^2\mathfrak{z}}+2\mathfrak{z}+\frac{2\mathfrak{z}'}{\mathfrak{z}}-\frac{6}{r}\right]
=\frac{2}{\mathfrak{z}}\left[\frac{1}{r^2}-8\pi\Pi\right].
\end{align}
This is a first order partial differential equation in $\mathfrak{a}$. Its solution after substituting back the definition of $\mathfrak{a}$ gives
\begin{align}\label{h81}
e^{\lambda(r)}=\frac{\mathfrak{z}^2
e^{\int\left(2\mathfrak{z}+\frac{4}{\mathfrak{z}r^2}\right)dr}}{r^6\left[2\int\left\{ \mathfrak{z}\left(\frac{1-8\pi\Pi r^2}{r^8}\right)
e^{\int\left(2\mathfrak{z}+\frac{4}{\mathfrak{z}r^2}\right)dr}\right\}dr+\bar{C}_1\right]}.
\end{align}
On can notice that the above model is presented in the form of two generating functions (GF), i.e., $\mathfrak{z}$ and $\Pi$. The corresponding matter variables for the locally anisotropic HSS become
\begin{align}\label{h82}
4\pi|\mu|&=\frac{m'}{r^2}-\Lambda,\\\label{h83} 4\pi
P_r&=\frac{\mathfrak{z}r(2m-r)-m+r}{r^3}+\Lambda,\\\label{h84} 8\pi
P_\bot&=\left(\frac{6m-3r+\Lambda r^3}{3r}\right)\left[\mathfrak{z}'
+\frac{1}{r^2}+\mathfrak{z}^2-\frac{\mathfrak{z}}{r}\right]+\mathfrak{z}\left[\frac{m'}{r}-\frac{m}{r^2}\right]+\Lambda.
\end{align}
In the following subsection, we shall describe families of GR-$\Lambda$ models for HSS under some realistic backgrounds.

\subsection{Conformally Flatness in HSS}

When a Riemannian space is conformally related to a flat Riemannian space, it is called \emph{conformally flat spacetime}. Thus for a conformally flat spacetime, the Weyl tensor vanishes equivalently, every point has a neighborhood conformal to an open subset of the Minkowski spacetime. As a result, a conformally flat spacetime owns the local conformal symmetry of the Minkowski spacetime, i.e., it has 15 independent conformal Killing vector fields. However, the global topology may not be the same as that of the Minkowski spacetime. In order to present
solution of HSS in this background, we take $\mathcal{E}=0$. Thus, Eq.\eqref{h41} becomes
\begin{equation}\label{h86}
\frac{1}{2}\frac{\partial}{\partial r}\left\{\frac{\lambda' e^{{-\nu}}}{r}\right\}+\frac{e^{-\nu-\lambda}}{2} \frac{\partial}{\partial r}\left\{\frac{\lambda' e^\lambda}{r}\right\}
-\frac{\partial}{\partial r}\left\{\frac{1+e^{-\nu}}{r^2}\right\}=0,
\end{equation}
which after substituting $\mathfrak{a}\equiv e^{-\lambda}$ and $\frac{\nu'}{2}\equiv \frac{\omega'}{\omega}$ make Eq.\eqref{h86} as
\begin{equation}\nonumber
\mathfrak{a}'+2\left(\frac{\omega''-\frac{\omega'}{r}+\frac{\omega}{r^2}}{\omega'-\frac{\omega}{r}}\right)\mathfrak{a}+\frac{2\omega}
{\left(\omega'-\frac{\omega}{r}\right)r^2}=0.
\end{equation}
This is again a first order differential equation, which gives
\begin{equation}\label{h89}
\mathfrak{a}=e^{-\int \xi(r)dr}\left(\int e^{-\int \xi(r)dr}k(r)dr+\mathcal{C}_4\right).
\end{equation}
After back substitution, we get
\begin{equation}\label{h92}
\frac{\lambda'}{2}-\frac{1}{r}=\frac{e^{\nu/2}}{r}\sqrt{-1+r^2\alpha_1 e^{-\lambda}}.
\end{equation}
In Eq.\eqref{h89}, $\xi$ and $k$ are defined as
\begin{eqnarray}\nonumber
&&\xi(r)=2 \frac{d}{dr}\left[ln\left(\omega'-\frac{\omega}{r}\right)\right],\\\nonumber
&&k(r)=\frac{-2\omega} {\left(\omega'-\frac{\omega}{r}\right)r^2},
\end{eqnarray}
while $\mathcal{C}_4$ is an integration constant, and $\alpha_1$ appearing in Eq.\eqref{h92} is also an integration function. Its value after matching
the interior region with de-Sitter spacetime provides
\begin{equation}\nonumber
\alpha_1=\frac{M(9M-4r)+\Lambda r^4/3}{r^4}.
\end{equation}
Equation \eqref{h92} gives
\begin{equation}\nonumber
e^\lambda=\alpha_1 r^2 \sin^2\left(\int\frac{e^{\nu/2}}{r}dr+\zeta\right),
\end{equation}
where
\begin{align}\nonumber
\zeta&\overset{\Omega^{(i)}}{=}\sin^{-1}\left[r\left\{\frac{\frac{2M}{r}-1+
\frac{\Lambda r^2}{3}}{M(9M-4r)+\Lambda r^4/3}
\right\}^{1/2}\right]-\left[\int\frac{e^{\nu/2}}{r}dr\right].
\end{align}
To proceed forward our solutions, we take another constraint, i.e., $P_r=0$, thereby considering the only one non-zero component of pressure in the gravitating source of HSS. Under this background, Eq.\eqref{h28} eventually provides
\begin{equation}\label{h96}
\lambda'=-\frac{1}{r}\left[1+e^\nu\left(1+\frac{\Lambda r^2}{8\pi}\right)\right].
\end{equation}
Upon using Eq.\eqref{h96} in Eq.\eqref{h41}, the conformal flat model of our system (after considering $e^{-\lambda}=2g-1$) can be recast as
\begin{equation}\label{h98}
g\left(9g+\frac{r^2\Lambda}{8\pi}-4\right)-rg'\left(3g-\frac{r^2\Lambda}{16\pi}-2\right)+\left(\frac{r^2\Lambda}{16\pi}\right)^2=0.
\end{equation}
This is again a first order differential equation. Its solution can easily be calculated as under
\begin{equation}\nonumber
g=\int \frac{4g\left(9g+\frac{r^2\Lambda}{8\pi}-4\right)+\left(\frac{r^2\Lambda}{16\pi}\right)^2}{4r\left(3g-\frac{r^2\Lambda}{16\pi}-2\right)}dr+\mathcal{C},
\end{equation}
where $\mathcal{C}$ is an integration function. By simultaneous making use of Eqs.\eqref{h92} and \eqref{h96}, one can find
\begin{equation}\nonumber
e^\lambda=\frac{\alpha_1r^2(2g-1)}{\left(3g+\frac{r^2\Lambda}{8\pi}-1\right)^2+2g-1}.
\end{equation}
In this way, we have been able to find unknown metric variable. Substituting this in the corresponding Einstein-$\Lambda$ equations of motion, we get structural variables as
\begin{align}\label{h101}
|\mu|=\frac{24g(2g-1)+\frac{r^4\Lambda}{\kappa^2}+\frac{2r^2\Lambda}{\kappa}}{16\pi r^2\left(3g-\frac{r^2\Lambda}{16\pi}-2\right)}-\frac{\Lambda}{\kappa},\quad P_\bot=\frac{24g^2+\frac{r^4\Lambda}{\kappa^2}+\frac{2r^2\Lambda}{\kappa}}{32\pi r^2\left(3g-\frac{r^2\Lambda}{16\pi}-2\right)}+\frac{\Lambda}{\kappa}.
\end{align}
The associated GF are calculated as under
\begin{align}\nonumber
z(r)=\frac{\Lambda r^2-g+1}{r(1-2g)},\quad \Pi(r)=-\frac{24g^2+\frac{r^4\Lambda}{\kappa^2}+\frac{2r^2\Lambda}{\kappa}}{32\pi r^2\left(3g-\frac{r^2\Lambda}{16\pi}-2\right)}-\frac{\Lambda}{\kappa}.
\end{align}
One can notice that the constraint $g-2/3>0$ on the metric coefficient $e^\lambda$ will make it to lie in the positive region. This also distributes the system to occupy the minimum value to
the radial coordinate. Thus the vacuum core can be expected to present within this HSS at the hypersurface $r=r_{min}$ even in the presence of cosmological constant. It is worthy to mention that this model describes the static solutions of conformally flat HSS which is assumed to be coupled with matter having
only one non-zero component of pressure. This model could be considered as a toy model for irrotational HSS cosmic bodies experiencing zero tidal forces due to $\mathcal{E}=0$.

\subsection{Non-Complex HSS Model}

Not only for static but also in non-static case, the structure scalar $Y_{TF}$ has been identified as the complexity factor in GR \cite{herrera2018new,herrera2018definition}. This result has also been found applicable in various modified theories of gravity \cite{yousaf2020complexity,bhatti2021structure}. One can study the non-complex Bondi as well as axially symmetric solutions by keeping $Y_{TF}=0$ \cite{herrera2019complexitya,herrera2019complexity}. It is easy to understand that such solutions are many in number even in the presence of cosmological constant. In order to specify them in a particular area field, we take a constraint, i.e., $P_r=0$. Thus Eq.\eqref{h28} yields
\begin{equation}\label{h126}
\lambda'=\frac{-2}{r(2g-1)}\left(g-\frac{\Lambda r}{8\pi}\right),
\end{equation}
with $g$ is same as mentioned before in the previous subsection. The zero complexity condition, i.e., $Y_{TF}=0$ gives
\begin{equation}\label{h128}
m_T\overset{\Omega^{(e)}}{=}(m_T)\left(\frac{r}{r}\right)^3.
\end{equation}
Equations \eqref{h46}, \eqref{h126} and \eqref{h128} provide
\begin{align}\nonumber
e^\lambda&\overset{\Omega^{(e)}}{=}\frac{4m_T^2(2g-1)r^4}{r^6\left(g+\frac{\Lambda r^2}{16\pi}\right)^2(\cosh\pi -1)^2}.
\end{align}
For the non-complex HSS isotropic matter configurations, we get
\begin{equation}\nonumber
rg'\left(\frac{\Lambda r^3}{32\pi}-g+1\right)+g\left(\frac{\Lambda r^2}{8\pi}-2+5g\right)-\frac{r^4\Lambda^2}{32 \pi^2}=0,
\end{equation}
which further gives
\begin{equation}\nonumber
\mathcal{C}_1=g-\int\frac{\left(\frac{r^2\Lambda}{\kappa}\right)^2-g\left(5g+\frac{\Lambda r^2}{8\pi}-2\right)}{r\left(\frac{\Lambda r^3}{32\pi}-g+1\right)}dr,
\end{equation}
thereby providing the value of $g$ by means of an integration function $\mathcal{C}_1$. The structural quantities for relativistic HSS are found as under
\begin{align}\label{h132}
|\mu|=\frac{3g\left(\frac{r^2\Lambda}{48 \pi}+2g-1\right)+\left(\frac{r^2\Lambda}{4\pi}\right)^2.\frac{1}{4}}{4\pi r^2\left(\frac{r^2\Lambda}{16\pi}+g-1\right)}-\frac{\Lambda}{\kappa},\quad P_\bot=\frac{g\left[3g\left(\frac{3r^2\Lambda}{32 \pi}+g-1\right)+\left(\frac{r^2\Lambda}{4\pi}\right)^2.\frac{1}{4}\right]}{8\pi r^2(2g-1)\left(\frac{r^2\Lambda}{16\pi}+g-1\right)}+\frac{\Lambda}{\kappa}.
\end{align}
In this environment, the GF are computed as
\begin{align}\nonumber
\mathfrak{z}=\frac{\Lambda r^2-g+1}{r(1-2g)},\quad \Pi=-\frac{g\left(\frac{3r^2\Lambda}{32 \pi}+g-1\right)+\left(\frac{r^2\Lambda}{4\pi}\right)^2.\frac{1}{4}}{8\pi r^2(2g-1)\left(\frac{r^2\Lambda}{16\pi}+g-1\right)}-\frac{\Lambda}{\kappa}.
\end{align}

\subsection{Stiff Fluid Configurations}

In this subsection, we shall provide the solution of HSS obeying a stiff state equation. This equations states that the difference between pressure and energy density should be equal to zero. Thus indicating it as $|\mu|=P_r$. With this setup, Eq.\eqref{h36} reads
\begin{equation}\label{h137}
P_r'=-\frac{2\Pi}{r}.
\end{equation}
This conservation equation could be applicable to those ultradense matter content which are distributing over the region by maintaining the constraints of stiff state equation. We proceed our analysis with the two assumptions. In the first case, we take $P_\bot=0$ and the second one keeps non-complex state of HSS. Thus, the first condition reduces Eq.\eqref{h137} to \begin{align}\nonumber
P_r=\frac{\mathcal{C}_3}{r^2}, \quad\quad
\Rightarrow|\mu|=\frac{\mathcal{C}_3}{r^2},
\end{align}
where $\mathcal{C}_3$ is an integration constant. In this context, the mass function and metric variables of HSS are found as under
\begin{align}\nonumber
m=\Lambda r^3+4\pi \mathcal{C}_3r,\quad e^{-\lambda}=2\Lambda r^2-1+8\pi \mathcal{C}_3,\quad\quad \nu=constant.
\end{align}
Then the associated GF are calculated as
\begin{equation}\nonumber
\mathfrak{z}=\frac{1}{r},\quad \Pi=\frac{\mathcal{C}_3}{r^2}.
\end{equation}
Now, we consider $Y_{TF}=0$. With this environment, the non-complex phase of HSS coupled with the matter obeying stiff sate equation can be observed. Equations \eqref{h74} and \eqref{h137} give
\begin{equation}\nonumber
P_r''+\frac{3P_r'}{r}=0.
\end{equation}
This is a first order differential equation whose solution is an easy task. It provides
\begin{equation}\label{h142}
P_r=\frac{\mathcal{D}_1}{r^2}-\mathcal{D}_2,
\end{equation}
where $\mathcal{D}_i's$ are constants of integration. The corresponding matter quantity for HSS is
\begin{equation}\nonumber
m=4\pi r\left(\mathcal{D}_1+\frac{\Lambda r^2}{3}-\frac{\mathcal{D}_2 r^2}{3}\right).
\end{equation}
With the help of this equation, one can get the values of metric co-efficient from Eq.\eqref{h35}. Since in HSS, the formation of Minkowskian core appears to emerge. This gives rise to the formation of two boundaries. At the outer hypersurface, i.e., $\Omega^e$, the associated values of radial component of pressure and mass function become
\begin{equation}\nonumber
P_r=\mathcal{D}_1\left[\frac{1}{r^2}-\frac{1}{r^2_{\Omega^e}}\right]-\Lambda,\quad
m=\frac{4\pi
\mathcal{D}_1r}{r^2_{\Omega^e}}\left(r^2_{\Omega^e}-\frac{r^2}{3}\right)
-\frac{4\pi}{3}\Lambda r^3+\frac{1}{3}\Lambda r^3.
\end{equation}
The difference between these expression provides
\begin{equation}\nonumber
4\pi P_r r^3-m=-\frac{16\pi}{3}\Lambda r^3+\frac{1}{3}\Lambda r^3+\frac{-8\pi \mathcal{D}_1r^3}{3r^2_{\Omega e}}.
\end{equation}
Eventually $P_\bot$ for the less complex HSS becomes
\begin{equation}\nonumber
P_\perp= -\frac{\mathcal{D}_1}{r^2_{\Omega e}}-\Lambda.
\end{equation}
These solutions are presented in the presence of cosmological constant.

\section{Energy Conditions}

\begin{figure} \centering
\epsfig{file=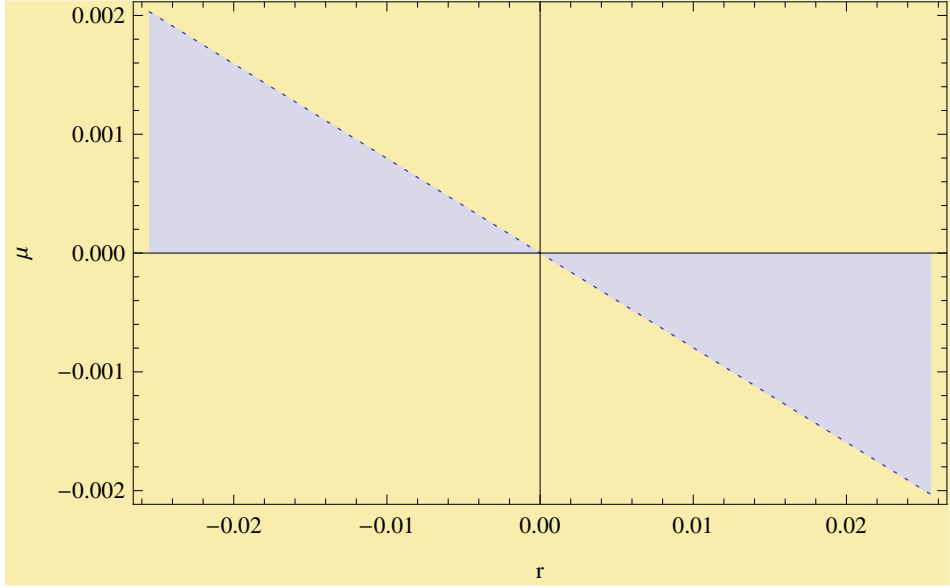,width=0.7\linewidth}
\caption{Behavior of energy density with the radial coordinate $r$. \label{energy1}}
\end{figure}

\begin{figure} \centering
\epsfig{file=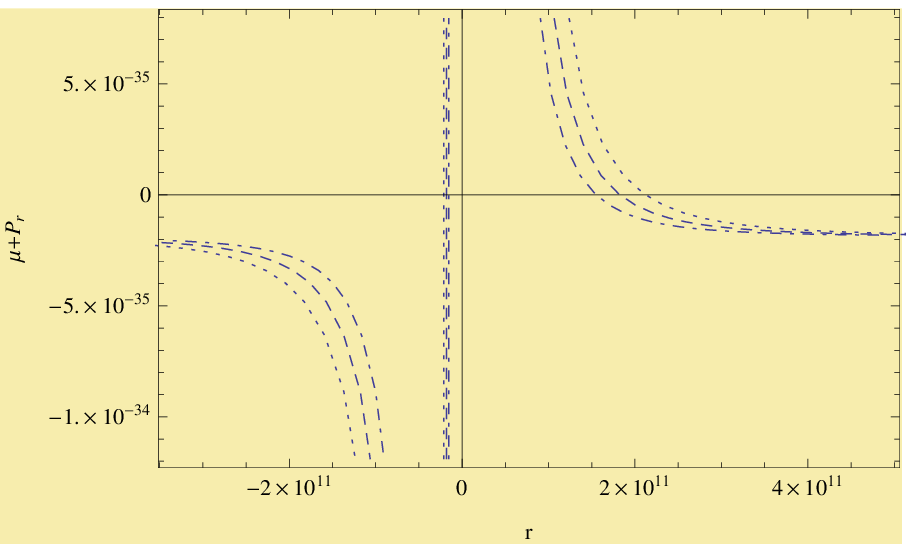,width=0.4\linewidth}
\epsfig{file=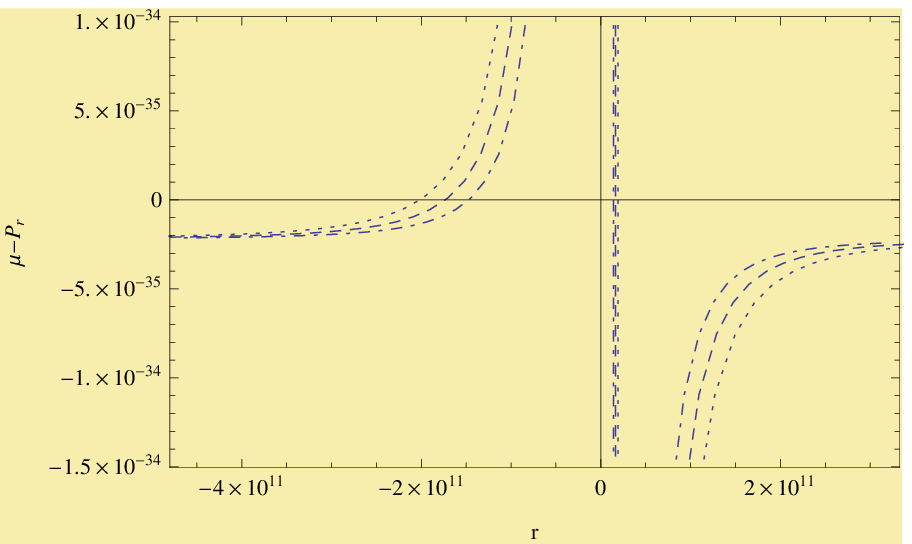,width=0.4\linewidth}
\caption{Analysis on WEC and DEC for three different strange stars. We represent the analysis on SS1, SS2 and SS3 with the dotdashed, dashed and dotted lines, respectively.   \label{energy2}}
\end{figure}

\begin{figure} \centering
\epsfig{file=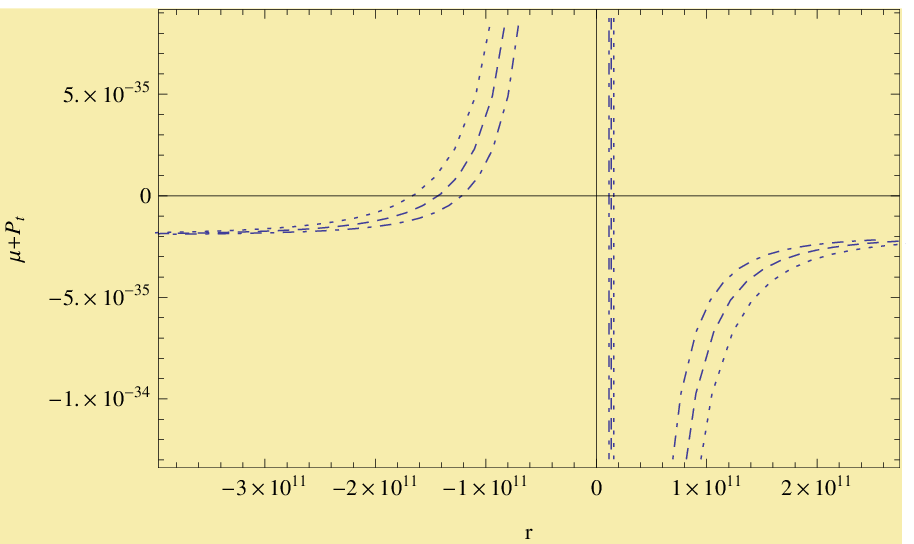,width=0.4\linewidth}
\epsfig{file=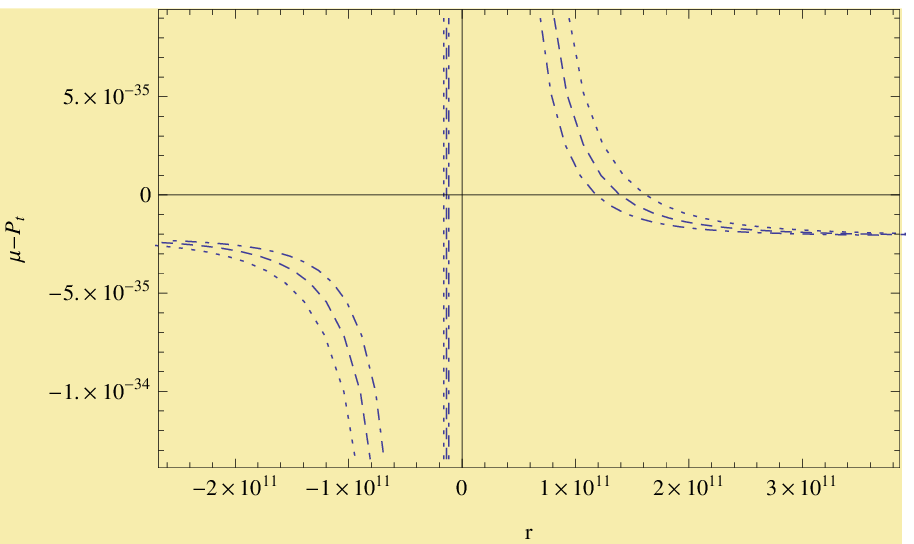,width=0.4\linewidth}
\caption{Analysis on WEC and DEC for three different strange stars. We represent the analysis on SS1, SS2 and SS3 with the dotdashed, dashed and dotted lines, respectively. \label{energy3}}
\end{figure}

\begin{figure} \centering
\epsfig{file=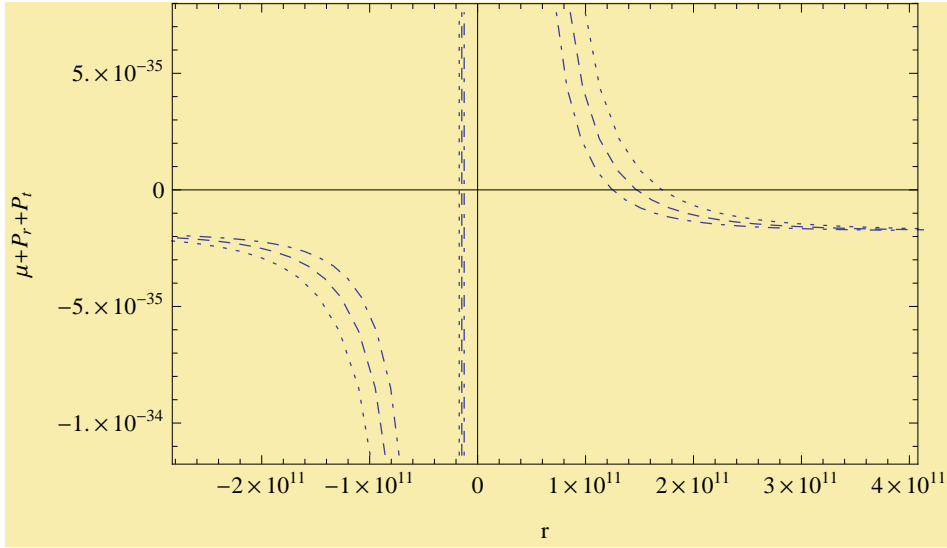,width=0.7\linewidth}
\caption{Analysis on SEC and DEC for three different strange stars. We represent the analysis on SS1, SS2 and SS3 with the dotdashed, dashed and dotted lines, respectively. \label{energy4}}
\end{figure}

The general form of the energy conditions can be obtained from the Raychaudhuri equation for expansion \cite{santos2007energy}. One can analyze the nature of gravity (attractive/non-attractive) from these conditions. The null energy conditions (NEC), weak energy conditions (WEC), strong energy conditions (SEC) and dominant energy condition (DEC) can be defined as follows
\begin{itemize}
  \item NEC : ${\rho + p_r} \ge 0$,  ${\rho + p_t} \ge 0$
  \item WEC : $\rho \ge 0$, ${\rho + p_r} \ge 0$,    ${\rho + p_t} \ge 0$
  \item SEC : ${\rho + p_r} \ge 0$,  ${\rho + p_t} \ge 0$,   ${\rho + p_r + 2p_t} \ge 0$
  \item DEC : $\rho \ge |{p_r}|$,    $\rho \ge |{p_t}|$
\end{itemize}
Figure \textbf{\ref{energy1}} indicates that the energy density of the hyperbolically symmetric static solutions is negative, if we take $\Lambda$ to be as a positive quantity. This suggests
the close connection of our results with quantum field theory. Such kind of analysis could be helpful in the examination of virtual particles, squeezed vacuum particle states and Casimir effect. Next we consider three different strange star candidates, i.e., Her X-1, SAX J 1808.4-3658, 4U1820-30. We would label these strange stars with SS1, SS2, and SS3, respectively. It is observationally seen that SS1, SS2, and SS3 have 0.88$M_{\odot}$, 1.435$M_{\odot}$ and 2.25$M_{\odot}$ stellar masses, respectively \cite{gangopadhyay2013strange,das2016compact}. We shall graphically observe the behavior of energy conditions for the toy model mentioned in Eq.\eqref{h81}-\eqref{h84} with respect to to stiff HSS generation functions. We have noticed from Figs. \textbf{\ref{energy2}}, \textbf{\ref{energy3}} and \textbf{\ref{energy4}} that the energy conditions are violated for majority of the zonal parameters. Only a few region is observed that support viability of energy conditions. It is worthy to mention that the observable effects of dark energy are well-known to violate the strong energy conditions. Therefore, our results could be helpful to understand dark energy problem as well as quantum or cosmological scenarios \cite{rubakov2014null,xiong2007violation,lasukov2020violation}.\\

\section{Conclusion}

The present paper is aimed to understand the modeling of hyperbolically symmetric solutions in the presence of cosmological constant. We have considered spatially hyperbolical spacetime which is assumed to be coupled with locally anisotropic fluid configurations. Under co-moving frame of reference and cosmological constant, we have presented field equations and Bianchi identities. With the help of two formalism (i.e., Misner-Sharp and Tolman) for the calculation of mass function, we have computed theoretical relations between total quantity matter and structural variables. The orthogonal splitting is used to decompose the curvature tensor into into scalar variables. Four structure scalars are calculated. This led us to understand the Tolman mass through $Y_{TF}$ and $Y_{T}$.

Finally, we proposed a generic framework that allows every diagonally symmetric hyperbolical fluid model with static backgrounds can be described by means of two generating functions. We have described the stellar models under three main categories. The first category gives cosmic solutions of hyperbollically symmetric spacetime in conformally flat backgrounds. The second describes the diagonally symmetric hyperbolical solutions for non-complex relativistic interiors. The last category is devoted to understand the structure formation through equation of state. Thus, certain straightforward corresponding solutions are discovered, along with with their physical explanations. It is necessary to address that the violation of the weak energy condition ($\mu<0$) in our modeling represents the negative attitude of Tolman mass under certain circumstances of cosmological constant and matter variables, i.e., if $4P_r r^3- m<\Lambda r^3$. Despite the fact that we would anticipate the energy density to be positive based on classical physics considerations. However, the negative energy densities are frequently mentioned in extreme astrophysical and cosmic events, notably in connection with quantum issues that could occur inside the horizon with cosmological constant.

The second order set of partial differential equations were solved in order to describe various physical properties of the hyperbolically symmetric objects in GR by Herrera et al. \cite{herrera2021hyperbolically}. The leading correction terms obtained in the present work are embodied with $\Lambda r^2$ terms stating that our solutions are depending directly on the impact of generating functions, $r$ and on the cosmological constant. This term could be interpreted as the non-attractive connection among the matter variables of the locally anisotropic HSS. The cosmological constant or vacuum energy density affects the impact of matter variables, introducing corrections to the results produced by Herrera et al. \cite{herrera2021hyperbolically} which could be of interest for situations describing in quantum and cosmological scenarios \cite{rubakov2014null,xiong2007violation,lasukov2020violation}. Corrections introduced by the cosmological constant produce variations on the metric coefficients as well as generating function due to their
repulsive nature for all the three fluid models.

We see that there exists some exact analytical solutions for the anisotropic HSS fluid models with the introduction
of a  $\Lambda$ term. Our study allows one to understand the implications of a nonzero $\Lambda$ towards the mathematical modeling of conformally flat, non-complex as well as stiff matter HSS distributions. These results are of interest in view of the recent observational claims about a non-vanishing cosmological constant. The other consequence of introducing the cosmological constant, however, concerns the mass definition. Herrera et al. \cite{herrera2021hyperbolically} described the matter content within the HSS through Misner-Sharp and the Tolman mass functions. Irrespective of the inclusion of the value of $\Lambda$ term, the positive and negative nature of the former and latter masses is never disturbed.


\section*{Acknowledgments}
{The present work is devoted to Prof. Dr. Muhammad Sharif (https://en.wikipedia.org/wiki/Muhammad\_Sharif \_(cosmologist)) on his 60th birthday.}

\section*{Conflicts of Interest}

{The author declares no conflict of interest.}

\end{document}